# Discovery of carbon-based strongest and hardest amorphous material

Shuangshuang Zhang[1†], Zihe Li[1†], Kun Luo[1,2†], Julong He[1†], Yufei Gao[1,2†], Alexander V. Soldatov[3,4,5], Vicente Benavides[3,6], Kaiyuan Shi[7], Anmin Nie[1], Bin Zhang[1], Wentao Hu[1], Mengdong Ma[1], Yong Liu[2], Bin Wen[1], Guoying Gao[1], Bing Liu[1], Yang Zhang[1,2], Yu Shu[1], Dongli Yu[1], Xiang-Feng Zhou[1], Zhisheng Zhao[1]*, Bo Xu[1], Lei Su[7], Guoqiang Yang[7], Olga P. Chernogorova[8], Yongjun Tian[1]*

[1]Center for High Pressure Science (CHiPS), State Key Laboratory of Metastable Materials Science and Technology, Yanshan University, Qinhuangdao, Hebei 066004, China

[2]Hebei Key Laboratory of Microstructural Material Physics, School of Science, Yanshan University, Qinhuangdao 066004, China

[3]Department of Engineering Sciences and Mathematics, Luleå University of Technology, SE-97187 Luleå, Sweden

[4]Department of Physics, Harvard University, Cambridge, MA 02138, USA

[5]Center for High Pressure Science and Technology Advanced Research, Shanghai 201203, China

[6]Department of Materials Science, Saarland University, Campus D3.3, D-66123, Saarbrücken, Germany

[7]Key Laboratory of Photochemistry, Institute of Chemistry, University of Chinese Academy of Sciences, Chinese Academy of Sciences, Beijing, 100190, China

[8]Baikov Institute of Metallurgy and Materials Science, Moscow 119334, Russia

* Corresponding authors: zzhao@ysu.edu.cn (Z.Z.) or fhcl@ysu.edu.cn (Y.T.). †These authors contributed equally to this work.

## ABSTRACT

Carbon is likely the most fascinating element because of the diversity of its allotropes stemming from its variable ($sp$, $sp^2$, and $sp^3$) bonding motifs. Exploration of new carbon forms has been an eternal theme of contemporary scientific research. Here we report on novel amorphous (AM) carbon materials containing high fraction of $sp^3$-bonded atoms recovered after compressing fullerene $C_{60}$ to previously unexplored high pressure and high temperature. Analysis of photoluminescence and absorption spectra demonstrates that they are semiconductors with tunable bandgaps in the range of 1.5-2.2 eV, comparable to that of amorphous silicon. The synthesized AM-III carbon is the hardest and strongest amorphous material known to date, capable of scratching diamond crystal and approaching its strength which is evidenced by complimentary mechanical tests. The remarkable combination of outstanding mechanical and electronic properties makes these AM carbon materials an excellent candidate for photovoltaic applications demanding ultrahigh strength and wear resistance. (148 words)

INTRODUCTION

Contrary to the crystalline state of solid matter which is characterized by periodicity in the spatial organization of the constituting atoms, the amorphous state exhibits no long-range order in the atomic arrangement although certain well-defined structural motifs may be present over a few interatomic distances giving rising to a degree of short- to medium-range order . The length scale over which such localized ordering occurs determines the physical properties for such systems. Another example is orientational disorder of molecules perfectly positionally arranged in a crystal. In both cases a common definition of the structure of these systems is disorder (spatial and/or orientational), which is also termed "glassy" state. Importantly, disordered systems exhibit many properties superior to their crystalline counterparts which makes them better candidates for technological applications. Bulk metallic glasses (BMG) have physical properties combining the advantage of common metals and glasses - strength several times higher than corresponding crystalline metals, good ductility and corrosion resistance [1–3]; hydrogenated amorphous silicon (a-Si:H) films exhibiting an optical absorption edge at ~1.7 eV have been the most popular photovoltaic semiconductor used in solar cells [4], and the a-Si:H/crystalline silicon (c-Si) heterojunction-based solar cell has increased efficiency steadily to a current record value of 24.7% [5], to name just a few examples. However, theoretical modeling of amorphous state is prohibitively difficult, thus exploring new amorphous states of matter and their nature is both rewarding and, at the same time, a very challenging scientific task of contemporary materials science.

Amorphous carbon exhibits a rich variety of physical properties determined by the ($sp$-$sp^2$-$sp^3$) bonding character and structural motif of the constituting atoms. Graphite-like $sp^2$ carbon, for example, is conductive, highly compressible and flexible due to disordered stacking of graphene layers in clusters. On the contrary, $sp^3$ bonding-dominated diamond-like carbon (DLC) films prepared by different deposition techniques from a large variety of carbon-carrying precursors exhibit high hardness, chemical inertness, and tunable optical bandgaps and, therefore, are widely used as protective coatings [6–8]. However, very large intrinsic stresses of up to several GPa in DLC films may result in the delamination of thick films from the substrates, and thereby limit the application of DLC coatings [9, 10].

Amorphous carbon can be alternatively synthesized by compression of $sp^2$ carbon precursors, typically fullerenes and glassy carbon (GC) [11–19]. Although $C_{60}$ molecules sustain pressure up to 20-25 GPa at ambient temperature [20], the buckyballs get easily broken already at about 5 GPa and elevated temperatures (~800 °C) to form a disordered nano-clustered graphene-based hard phase with more than 90% elastic recovery after deformation [21, 22]. Likewise, disordered carbon materials with different $sp^2$-$sp^3$ carbons ratios exhibiting a remarkable combination of lightweight, high strength and elasticity together with high hardness and electro-conductivity can be recovered after compressing GC at pressures of 10-25 GPa and high temperatures ≤ 1200 °C [11]. With further increase of pressure the GC transforms into a metastable $sp^3$-rich dense, ultra-incompressible amorphous carbon [12–14]. Importantly, synthesis of carbon allotrope capable of scratching diamond by exposure of fullerene $C_{60}$ to 13 GPa, 1227-1477 °C with subsequent quenching to ambient conditions has been reported [17], although properties of this phase and interpretation of its structure remain a subject of unresolved controversy. Even though the great efforts have already been put into exploration of the p,T phase diagram of $C_{60}$, the pressure range above 20 GPa, has yet to be established. As synthesis pressure strongly affects the microstructure and bonding in carbon phases produced from $C_{60}$, we may envisage emergence of new amorphous carbon polymorphs as a result of crystal-to-amorphous and/or amorphous-to-amorphous phase transitions triggered in the pressure range of the structural integrity of $C_{60}$ [23, 24].

Here, we present a systematic study of the behavior of $C_{60}$ fullerene at previously unexplored pressure of 25 GPa and different temperatures. Amorphous carbon materials, namely AM-I, AM-II and AM-III, were synthesized and characterized by complimentary techniques: X-ray diffraction (XRD), Raman spectroscopy, high-resolution transmission electron microscopy (HRTEM), and electron energy loss spectroscopy (EELS). Our results demonstrate that the $sp^3$ carbon fraction in these materials gradually increases with increase of the synthesis temperature and, finally, reaches 69%-94%. Different hardness measurement methods including Knoop ($H_K$), Vickers ($H_V$) and nanoindentation hardness ($H_N$) together with uniaxial compressive strength test were employed in order to assure the reliability of the obtained results and demonstrate that the synthesized AM-III *bulk* material is the hardest and strongest amorphous material known to date. In addition, unlike insulator diamond, these AM carbon materials are semiconducting with relatively narrow bandgaps (1.5-2.2 eV) that opens up good perspectives for using these materials in new class of photoelectric applications.

**RESULTS AND DISCUSSION**

**Structural characterization**

Fig. 1A and B show XRD patterns of the materials recovered after treatment of $C_{60}$ at 25 GPa and various synthesis temperatures. The following sequence of phase transitions was observed: first, $C_{60}$ transforms into the known 3D polymer [17] at elevated temperature (new sharp diffraction peaks appear), then buckyballs destruction/structure amorphization begins at about 500 °C (very broad new peaks appear and the intensity of the polymer peaks decreases) and completes above 800 °C. The materials recovered from 1000 °C, 1100 °C and 1200 °C, termed as AM-I, AM-II and AM-III, respectively, are characterized by a dominant broad diffraction peak centered near q=~3.0 Å$^{-1}$, fairly close to the position of (111) reflection of diamond (q=3.05 Å$^{-1}$) and a weaker peak at q=~5.3 Å$^{-1}$ (Fig. 1A and Supplementary Fig. 1), which represent an entirely new class of amorphous carbon materials distinctly different from the previously reported low-density amorphous carbon materials synthesized at lower pressures and temperatures (13 GPa, 1227-1477 °C) [17]. Recently, Shi and Tanaka revealed that the first sharp diffraction peak (FSDP) in tetrahedral covalent amorphous materials such as Si, Ge and C comes from the characteristic density waves of a single tetrahedral unit, and the integrated intensity of FSDP is directly proportional to the fraction of locally favored tetrahedral structure or a measure of the tetrahedrality [25]. Notably, the previously discovered amorphous carbon materials have another graphite-like diffraction peak near q=~2.0 Å$^{-1}$ indicating the large interlayer spacing and lower density [18] (see also the results of our test experiment conducted at similar conditions as described in Ref. 19, Supplementary Fig. 2). With the synthesis temperature increase from 1000 °C to 1200 °C, the amorphous peaks become slightly narrower and shift from ~2.88 to ~3.00 Å$^{-1}$, indicating further density increase. Also, the material's color changes from opaque black to transparent yellow (insets in Fig. 1C). As the synthesis temperature exceeds 1300 °C, the narrow diffraction peaks corresponding to (111), (220) and (311) reflections of diamond appear near 3.05, 4.98 and 5.84 Å$^{-1}$, respectively, indicating the formation of nanocrystalline diamond (nano-diamond) coexisting with the remaining amorphous phase.

The bonding difference in the AM carbon materials is reflected in their Raman spectra (Fig. 1C and Supplementary Fig. 3). The AM-I and AM-II exhibit a broad Raman peak around 1600 cm$^{-1}$ with full width at half peak maximum (FWHM) of ~200 cm$^{-1}$, corresponding to G-band characteristic of *sp$^2$* carbons. Appearance of the G-band peak testifies for relatively high fraction of *sp$^2$* bonded carbon atoms [26]. Indeed, accounting for very low Raman cross-section for *sp$^2$* carbon at UV laser excitation, the high intensity of the G-band in the spectra of AM-I and AM-II clearly indicates on the *sp$^2$* carbon dominance in these amorphous materials. Importantly, both position and the FWHM of the G-band peak indicate the Raman scatterers' (clusters) size in these materials must be less than 2 nm [27]. On the contrary, the background-subtracted Raman spectrum of AM-III reveals several new features. First, a band located at the low wavenumbers of 900~1300 cm$^{-1}$ (termed as "T-band" [27]) is a characteristic signature of *sp$^3$* carbon and thus indicates on their high concentration in the AM-III. Second, an evident shoulder (rising peak) on high frequency side of the G-band (at 1740 cm$^{-1}$) may be attributed to clustering (cross-linking via *sp$^3$* bonds) of remaining aromatic rings formed of *sp$^2$* carbon and, finally, the peak appearing at about 1930 cm$^{-1}$ is likely originating from short linear chains (Supplementary Fig. 3B). After completion of the AM-diamond transformation above 1600 °C, the fingerprint peak of crystalline diamond at ~1330 cm$^{-1}$ appears in the spectra of transparent diamond samples (see inset in Fig. 1C).

In order to confirm the microstructure and bonding nature of the AM carbon materials suggested by Raman, HRTEM, selected area electron diffraction (SAED) and EELS were performed. The SAED patterns display two diffuse rings near 2.1 Å and 1.2 Å in all three AM carbon materials (Fig. 2), that is consistent with the XRD results. For comparison, the composite sample recovered from 1300 °C shows, in addition, the "spotty" diffraction rings indicating the formation of nanocrystalline diamond. The main feature of the low loss EELS data in Fig. 2C is a gradual shift of the plasmon peak from its position in pristine $C_{60}$ (26.0 eV) to higher energies in AM-I, AM-II and AM-III (29.7, 30.7, and 32.8 eV, respectively) that demonstrates increase of $sp^3$ fraction in the AM carbon materials. The plasmon peak position in AM-III is higher than that in the "amorphous diamond" (a-D) produced by quenching GC from high p,T (31.8 eV) that implies lower $sp^3$ content and density in the latter [12]. According to the plasmon peak position in the low loss EELS spectra (Supplementary Fig. 4A), the $sp^3$ fraction in AM-I, AM-II, and AM-III was estimated to be 69±4%, 77±2%, and 94±1%, respectively, similar to the method described previously [28]. Density of AM-I, AM-II and AM-III was directly measured at ~2.80±0.17, ~2.96±0.08, and ~3.30±0.08 g/cm$^3$, respectively, thus demonstrating AM-III is the densest amorphous carbon approaching diamond. The $sp^3$ fraction value was also independently determined based on the density of AM carbon materials using the calibration plot of $sp^3$ fraction vs. density [29], as shown in Supplementary Fig. 4B, which is similar to above results estimated from plasmon peak position. In addition, the peak at 285 eV in carbon K-edge (high loss) EELS signaling the $sp^2$ fraction in the material gradually decreases when going from AM-I to AM-III (Fig. 2F). The linear EELS scans with high spatial resolution in randomly selected sample regions demonstrate the bonding homogeneity at least on 1 nm scale in these AM carbon materials (Supplementary Fig. 5). The subtle microstructure differences between the AM carbon materials are further revealed by HRTEM images that exhibit a characteristic "worm-like" contrast manifesting structural disorder (Fig. 2A, B and D). The dimensions of these very fine structural fragments gradually decrease with the synthesis temperature increase, reaching statistically averaged size of about 12 Å, 8 Å and only 4 Å in AM-I, AM-II and AM-III, respectively. That clearly distinguishes these disordered carbon materials from those containing substantially lower fraction of $sp^3$-bonded atoms obtained from GC at similar p,T conditions [11], underscoring importance of the precursor material selection in high p,T synthesis.

**Mechanical properties**

The hardness values, i.e. $H_K$, $H_V$ and $H_N$, of the AM carbon materials were estimated by three independent measurement methods. The results as well as detailed indentation images are presented in Fig. 3 and Supplementary Figs. 6 and 7. Among the synthesized materials, AM-III has the highest hardness of $H_K$ =72±1.7 GPa and $H_V$ =113±3.3 GPa, whereas the AM-I and AM-II have $H_K$ of 58±1.9 and 62±1.9 GPa, respectively. In comparison, the $H_V$ and $H_K$ values of (111) plane of natural single crystalline diamond are 62 and 56 GPa [30, 31], respectively (Fig. 3A and Supplementary Fig. 8), thus hardness of the synthesized AM carbon materials can rival that of diamond. Careful analysis of Vickers indentation morphologies of AM-III shows that the raised "pile-up" was formed due to flow of the displaced material up around the indenter, indicating the plastic character of the deformation during loading (Supplementary Fig. 9C). With the applied load increase up to 3.92 N, the radial and lateral cracks as well as the peeling zone can be observed around the resultant indentations (Supplementary Fig. 9A and B), implying occurrence of the plastic-to-brittle transition in the material [32]. Moreover, the $H_N$ and Young's modulus ($E$) have also been determined based on the load-displacement curves using Oliver and Pharr model [33] (Supplementary Fig. 7). The estimated $E$ of AM-I, AM-II and AM-III are 747±66, 912±89, 1113±110 GPa, respectively. The obtained $H_N$ for them are 76±3.4, 90±7.9, and 103±2.3 GPa, respectively, which are comparable to their Vickers hardness. Notably, the $H_N$ of AM-III exceeds the record of 80.2 GPa held until now by tetrahedral amorphous carbon (ta-C) films [8]. Such extreme hardness allows the AM-III sample scratch the (001) face of synthetic diamond crystal with $H_V$ of 103 GPa (Fig. 3C and Supplementary Fig. 8A). Possessing hardness comparable to that of single crystalline diamond, AM-III becomes the hardest amorphous material known to date (Fig. 3B). More significantly, the advantage of these ultrahard amorphous carbon is that they have isotropic hardness comparable to diamond crystals where the hardness varies along different crystallographic directions leading to a cleavage of diamond easy to occur along its "weak" crystal planes.

The superior mechanical properties of AM-III have been further demonstrated by *in-situ* uniaxial compression/decompression test (Supplementary Fig. 10). It was found that micropillar made out of the AM-III with top diameter of 0.88 μm has compressive strength of at least 40 GPa, and could be fully elastically recovered without fracture after decompression at ambient conditions. Subsequent measurement of a micropillar with larger top diameter (3.78 μm) demonstrated its ability to withstand compressive stress as high as ~70 GPa without fracture although in this case a closer examination of the decompressed pillar revealed some wrinkles produced in its upper part (inset in Fig. 4A), very similar to the shear bands formed in metallic glasses during deformation [3]. Another AM-III micropillar with diameter of 2.64 μm was broken at stress load of 65 GPa before reaching its strength limit. Thus the measured compressive strength of AM-III lies in between that of <100>- and <111>-oriented diamond micropillar exhibiting the compressive strength of ~50 GPa and ~120 GPa, respectively [34]. Theoretically, the maximum compressive strength of materials can only be obtained when the compression direction is strictly parallel to the normal to the measured sample surface, the condition which is very difficult to achieve. As a result, the value of ideal compressive strength of the amorphous AM carbon pillars should, in fact, be higher than that we determined in our experiment. Consequently, our measurements demonstrate that the AM-III is comparable in strength to diamond and superior to the other known high-strength materials (Fig. 4B) [34–36].

It is important to ascertain what may be the reason(s) for the observed AM carbon materials with $sp^3$ carbon fraction below 100%, in particular AM-I with only 69% $sp^3$, exhibiting hardness and strength comparable to that of crystalline diamond. Indeed, it is well known, the $sp$, $sp^2$ and $sp^3$ covalent bonds in elemental carbon are all extremely strong. For example, the intrinsic strength of graphene (pure $sp^2$ carbon) reaches a value as high as 130 GPa [37] thus exceeding ultimate tensile strength of diamond <111> direction (95 GPa [38]) comprised of $sp^3$ carbons. The fundamental reason for the softness of graphite is weak van der Waals interaction between graphene layers. However, high pressure induces partial $sp^2$-to-$sp^3$ transformation leading to interlinking/locking-in the graphene layers by the tetrahedral $sp^3$ bonds and profound increase of hardness and strength of the resulting high-pressure phase that is able to abrade the diamond anvils [39]. Such as $sp^2$-$sp^3$ carbon system with only 22% $sp^3$ fraction experimentally obtained at ambient conditions by quenching from high-pressure compressed GC has high hardness of 26 GPa [11], whereas the three-dimensional (3D) $C_{60}$ polymer comprised of covalently linked (via $sp^3$ bonds) fullerene molecules with ~40% $sp^3$ carbons content, possesses superhigh hardness of 45 GPa [40]. Moreover, a number of superhard/ultrahard $sp^2$-$sp^3$ crystalline carbon forms were recently predicted theoretically. For example, the carbons designated as P-1-16b, P-1-16e, and P-1-16c with ~50% $sp^3$ carbons are predicted to have ultrahigh hardness of 71.3-72.4 GPa [41]. A series of superhard $sp^2$-$sp^3$ 3D carbon nanotube polymers such as the 3D (8,0) nanotube polymer with 43.5% $sp^3$ carbon is predicted to have superhigh hardness of 54.5 GPa [42, 43]. And a class of diamond-graphene (diaphite) nanocomposites constructed from covalently connected $sp^3$-diamond and $sp^2$-graphite structural units are predicted to high hardness and improve fracture toughness [44, 45]. All the above mentioned experimental and theoretical results demonstrate that ultrahigh hardness and strength comparable to crystalline diamond can be achieved in $sp^2$-$sp^3$ carbon systems at $sp^3$ concentrations below 100%. The AM carbon materials synthesized in this work have higher $sp^3$ contents than that in compressed GC [11] and 3D-$C_{60}$ polymers [40] thus we may anticipate higher hardness and strength in our systems. More importantly, it's not just a fraction of $sp^3$ carbon atoms that matters in this case but the structural motif. We argue that our $sp^2$-$sp^3$ carbon systems represent a particular short-range order which is a "blend" of remaining $sp^2$ carbon-based units (fused aromatic rings, short chains) covalently interlinked with clusters of tetragonally-coordinated $sp^3$ carbons. Such a "blend" represented on the HRTEM images (Fig. 2A, B and D) by a worm-like structural fragments must combine nearly intrinsic graphene-type strength/hardness of the $sp^2$ units with diamond-like strength/hardness of the clusters formed by tetragonally-coordinated $sp^3$ carbon. That may explain why already AM-I with relatively low $sp^3$ fraction is competitive in hardness and strength with crystalline diamond. On developing of substantially smaller structural fragments (fused rings opening, interlinking the structural units via short chains) along with significant increase of $sp^3$ fraction in AM-III, a new short-range order must emerge and further manifest in profound increase of hardness, strength and altering the optical properties of the system.

**Optical properties**

These AM carbon materials under investigation display also unusual optical properties. Through at a wavelength of 532 or 633 nm laser excitation, all the materials exhibit strong photoluminescence (PL) in range of 550-950 nm (Fig. 5A). The PL maxima correspond to photon energy of 1.59±0.1, 1.74±0.2 and 1.87±0.1 eV, in AM-I, AM-II, AM-III, respectively. This difference is directly related to the higher content of $sp^3$ carbon-based material possessing larger bandgaps in the samples. In view of yellow-transparent nature of AM-III, its visible light absorption spectrum was measured in transmission utilizing a diamond anvil cell (DAC). The inset of Fig. 5B shows the view in transmitted light through a sample piece mounted in a gasket hole inside the DAC. The result indicates that the optical absorption edge of AM-III is located at the ~570 nm, which corresponds to a bandgap of 2.15 eV, consistent with the PL results. Therefore, these AM carbon materials are a class of semiconductors with bandgaps less than diamond (5.5 eV) and close to amorphous silicon (a-Si:H) films (~1.7 eV) wildly used in technology nowadays. The preferable optical bandgaps offer a potential of using these AM carbon materials as optimal semiconductors for novel photoelectric applications.

**Comparison of various types of amorphous carbon**

It is important to define position of the materials we produced on the current landscape of other technologically important (hard) amorphous carbon-based materials. The data reported/published to date can be divided into 2 categories according to the preparation method: thin films prepared by various deposition routes [8, 27, 46, 47] and the materials synthesized at high-pressure and high-temperature using different precursors such as fullerene[12, 17] and GC [9-12]. *Further we mainly focus on the most distinct material – AM-III* (Fig. 6). Comparing the microstructure and bonding of the discovered AM-III to ta-C(:H) films [8, 27, 46, 47] through the correspondent UV Raman and EELS (Fig. 6A, D and E), one can see a much stronger Raman T-band around 900~1300 cm$^{-1}$ characteristic of *sp*$^3$ carbon and a negligible EELS intensity in the AM-III against the peak near 285 eV representing residual *sp*$^2$ carbon in ta-C(:H) films [46, 47]. Importantly, the residual *sp*$^2$ carbon presents in the films in the form of orientationally disordered nano-sized graphene clusters whereas no graphene-based structural units survive 25 GPa synthesis pressure in AM-III we report here. The evident structural difference results in significant performance difference between these materials. For example, the AM-III has a high $H_N$ of 103 GPa, which is comparable to the hardest crystal plane of diamond, and higher than that (80.2 GPa) of the reported "hardest" ta-C film [8].

In the second category, the hard amorphous carbon materials were produced at high p,T from fullerene and GC precursors with synthesis pressures up to 15 GPa [17] and 50 GPa [12], respectively. The XRD patterns in Fig. 6B exhibit clear difference between AM-III and various AM carbon materials synthesized previously by compressing $C_{60}$ at relatively low synthesis pressures (up to 15 GPa) [17] - the graphite-like diffraction peaks near q=~1.5-2.0 Å$^{-1}$ still appear in the XRD patterns indicating presence of large interlayer spacings and, consequently, relatively low densities. These highly-disordered *sp*$^2$ carbon-based systems exhibit graphene nanoclusters-derived short range-order that is preserved at the synthesis pressure used in earlier experiments, which is evidenced in both Raman and HRTEM data (Fig. 6A, and C) [17, 47]. In order to further reveal the characteristics of this type of AM carbon materials, we undertook a special effort and performed synthesis at p,T conditions (15 GPa, 550-1200 °C, see Supplementary Fig. 2) similar to those used in Ref. 19. Apparently the Vickers hardness of the material we synthesized at 15 GPa, 800 °C (see its HRTEM in Fig. 6C) was found to be 68 GPa, i.e. lower than that of newly synthesized AM carbon materials (Fig. 3A). Thus: i) testifying for presence of an entirely different type of short-range order and composition ($sp^2/sp^3$ ratio) in the system and ii) demonstrating that fullerene compression at a level of 25 GPa is an essential requirement to facilitate both altering the short-range order (crushing the residual nano-graphene clusters) and $sp^2$ to $sp^3$ transformation/formation of the tetragonal amorphous carbon matrix. On the contrary, using GC comprised of relatively large, irregular and curved multilayer graphene sheets as the precursor demonstrated that one must go to much higher pressure than 25 GPa in order to create $sp^3$ carbon-based material as graphene nanoclusters formed by crushing the curved graphene sheets in GC survive at this synthesis pressure and exhibit super-elastic properties when quenched to ambient conditions (see its HRTEM in Fig. 6F) [11]. Indeed, laser heating to ~1527 °C at 40-50 GPa allowed to produce a $sp^3$-rich system, so called "quenchable amorphous diamond" (a-D) [12]. It is important to underscore the big difference between microstructures of the $sp^3$ carbon-based AM carbon materials we synthesized from compressing $C_{60}$ and a-D [12] which is evident from HRTEM images - very high structural homogeneity with uniform and ultrafine structural units/fragments in the former (Fig. 2A, B, and D) and non-uniform, inhomogeneous contrast with larger size structural fragments overlapping with an additional contrast from crystalline planes with low spacing planes in the latter [12]. In addition, XRD pattern of a-D reveals the signature of a residual peak at about 2 Å$^{-1}$ corresponding to graphite-like interlayer distance and low EELS data indicate higher residual $sp^2$ carbon contents in a-D compared to that in AM-III (Fig. 1B). That underscores clear difference between these amorphous carbon forms. The comparison of amorphous produced from GC to the materials synthesized in this work demonstrate ultimate importance of the precursor material in the high p,T synthesis. Indeed, using highly symmetrical intrinsically nanostructured $C_{60}$ molecule (only ~7 Å in diameter) as a precursor provides uniform bonds breaking and conversion along with amorphization of the structure under 25 GPa, 1000-1200 °C compared to GC where even pressure increases to 50 GPa was insufficient to turn it at ~1527 °C into a uniform $sp^3$ carbon-based structure.

Thus, the AM carbon can be divided into at least five categories according to our understanding, by summarizing all the reported and currently synthesized materials. *The first type* is all-$sp^2$ disordered carbon materials composed of curved graphite-like or multilayer graphene fragments with variable sizes and microstructures (e.g., 5-, 6- or 7-membered rings), such as carbon black, GC and other AM carbon materials formed from high-temperature carbonization of organic compounds. *The second type* is mainly composed of curved multilayer graphene fragments with variable sizes and microstructures, and a small amount of $sp^3$ bonds formed between the layers of multilayer graphene fragments, such as compressed GC [11] and a-C(:H) films formed by deposition [6, 7]. For this kind of AM carbon materials, an obvious graphite-like diffraction of q=~1.5-2.0 Å$^{-1}$ can be observed (Fig. 6B, sample ①). Compared with *the first type* of all-$sp^2$ AM carbon, this kind of AM carbon materials have significantly improved mechanical properties such as high hardness/strength, but also has good conductivity due to the $sp^2$ bonding dominant. *The third type* is composed of $sp^3$ dominant dense disordered component and disordered nano-multilayer graphene fragments, such as the ta-C(:H) films [8, 27, 46, 47] and currently synthesized AM-I, AM-II, as well as the AM carbon materials recovered from compressing $C_{60}$ at 15 GPa and 800-1000 °C. For this kind of AM carbon materials, the diffraction peak from multilayer graphene in structure at q=~2.0 Å$^{-1}$ still exists, but becomes weak. At the same time, the intensity of diffraction peak at q=~3.0 Å$^{-1}$ from diamond-like tetrahedral structure gradually increases with the transformation and decrease of the multilayer graphene component in microstructure.

This type of AM carbon are semiconducting materials with superhigh hardness and strength. *The fourth type* is composed of $sp^3$ dominant dense disordered structure, such as AM-III currently synthesized. This type of AM carbon materials has no diffraction peak from the multilayer graphene interlayer (q=~1.5-2.0 Å$^{-1}$) and only has a broad diffraction peak centered at ~3.0 Å$^{-1}$, which is close to the position of (111) reflection of diamond. *The fifth type* is an ideal AM carbon with complete $sp^3$-bonded carbon atoms.

Going further we must underscore that contrary to crystalline materials where using just one technique, XRD, for example is sufficient for distinguishing different structural states, a complimentary characterization of the AM carbon materials is mandatory as it allows for clear identification of different states of disordered matter. Only using complimentary characterization comprised of XRD, Raman, HRTEM and EELS allowed us not only to distinguish the newly synthesized AM carbon materials from all other AM carbon materials reported to date but also to reveal subtle differences between these novel structural forms of carbon. For example, whereas the difference between AM-III and AM-I/AM-II is evident the latter materials are hard to distinguish when we look just at their Raman spectra (Fig. 1C and Supplementary Fig. 3). On the contrary, the EELS data indicate the difference in $sp^3$ fraction between all the AM carbon materials (Fig. 2C, F and Supplementary Fig. 4), and the HRTEM demonstrates the homogeneous contrast but distinct difference in the size of the structural worm-like fragments in the AM carbon materials (Fig. 2A, B and D). We infer that evolution from AM-I to AM-II state likely goes via relaxation of the structure around crushed buckyballs triggered by temperature increase at 25 GPa - fusion of the remaining aromatic rings built of $sp^2$ carbons, further carbon conversion from $sp^2$ to $sp^3$ state and bridging the fused rings and clusters of tetragonally-coordinated $sp^3$ atoms. A more profound change in the short-range order occurs in AM-III leading to the aromatic rings opening, short chains formation (evidenced by appearance of new Raman peak at 1940 cm$^{-1}$, Supplementary Fig. 3B) and accompanied by interlinking of the structural elements via $sp^3$ carbon which fraction substantially increases on this step. Consequently, these structural differences result in different performance of the AM carbon materials, in particular, mechanical and optical, properties as discussed in detail above.

The above analysis demonstrates that the discovered AM-III is indeed a new amorphous carbon material never detected and reported before. The distinct short-range order, microstructure and composition provide a unique combination of semiconducting and superior mechanical properties (with hardness and strength at the level of natural/synthetic diamond).

## CONCLUSION

In summary, by extending synthesis pressure to 25 GPa AM carbon material was created from $C_{60}$ precursor. Higher synthesis pressure seizes growth of graphene clusters after buckyballs collapse leading to high enrichment of the synthesized disordered phases with $sp^3$-bonded carbon, thus concluding the search for a bulk material based on tetragonally-arranged $sp^3$ carbon network finally complimenting and expanding technological value of the existing 2D systems – ta-C and DLC films. Consequently, the materials exhibit outstanding mechanical properties – comparable to crystalline diamond, and the hardness and strength of AM-III surpass any known amorphous materials. Thermal stability of AM-III in-air is comparable to that of diamond crystals [30] (Supplementary Fig. 11). Remarkably, these AM carbon materials are all semiconductors with the bandgaps in the range of 1.5-2.2 eV. The emergence of this type of ultrahard, ultrastrong, semiconducting AM carbon material offers excellent candidates to most demanding practical applications and calls-up for further experimental and theoretical exploration of the amorphous carbon allotropes.

## METHODS

### Sample synthesis

The samples with diameters of ~1 mm and heights of 1.2-1.7 mm were recovered after compressing $C_{60}$ fullerene (99.99%, Alfa Aesar) at pressure of 25 GPa at high temperatures. The standard COMPRES 8/3 sample assembly consisting of a 8-mm-spinel ($MgAl_2O_3$) octahedron with a Re heater and a $LaCrO_3$ thermal insulator was used for high-pressure (P ~25 GPa) and high-temperature (T ~2300 °C) experiments in a large-volume multi-anvil system at Yanshan

University, identical to the described elsewhere [30]. Pressure loading/unloading rates were 2 GPa/hour, heating rate was 20 °C/min, maintained for 2 hours and finally quenched by turning off the electric power supply.

**X-ray diffraction and Raman spectroscopy**

X-ray diffraction (XRD) was performed on a Bruker D8 Discover Advanced diffractometer with Cu *Kα* radiation source.

Both Raman scattering and photoluminescence (PL) measurements were carried out on a Horiba Jobin–Yvon LabRAM HR-Evolution Raman microscope at ambient conditions. The Raman spectra were excited by the laser radiation of 325 nm, and the PL spectra were excited by 532 or 633 nm laser. In all experiments the laser beams was focused to a spot size of ~1 μm.

**HRTEM and EELS measurements**

The samples for HRTEM were prepared by a Ga focused ion beam (FIB) (Scios, FEI) milling with an accelerating voltage of 30 kV. HRTEM, SAED, and EELS measurements were carried out at Themis Z TEM, using accelerating voltage of 300 kV. The EELS spectra were collected in the TEM mode at a random region of ~200 nm. The EELS line scans were conducted in STEM mode with an energy resolution of 0.6 eV and spatial resolution of ~1 nm.

**Hardness and elastic modulus measurement**

$H_K$ and $H_V$ were measured by microhardness tester (KB 5 BVZ), and the adopted loading and dwelling time were 40 s and 20 s, respectively. $H_N$ and Young's moduli ($E$) were measured at the peak load of 0.98 N with a three-sided pyramidal Berkovich diamond indenter (Keysight Nano Indenter G200). The indentations were imaged by the Atomic Force Microscope (AFM) to obtain the accurate hardness. The scratch test was conducted by using AM-III as an indenter to scratch the (001) crystal face of diamond.

**Compressive strength test**

The micropillars with diameters of ~1 to 4 μm and aspect ratios of ~1.5 to 2.5 were fabricated using a Ga ion beam at an accelerating voltage of 30 kV in FIB instrument (Scios, FEI). The compression measurements were conducted at PI 87 PicoIndenter system interfaced with a Helios NanoLab DualBeam microscope and nanoindentation system (Keysight Nano Indenter G200).

**Optical absorption**

The VIS/NIR absorption spectra were recorded on a UV/VIS/IR spectrometer (Avantes, AvaSpec) using a Xenon Light Source by assembling a sample piece with thickness of ~50 μm into a diamond-anvil cell (DAC) with a culet size of 500 μm. The band gaps were derived from the absorption spectra using the method described elsewhere [48].

**Thermal stability measurement**

Differential scanning calorimetry (DSC) and thermogravimetric analysis (TGA) using NETZSCH STA 449F5 were measured in the temperature range of 25-1400 °C with a heating rate of 10 °C/min.

**FUNDING**

This work is supported by National Science Foundation (NSF) of China (Nos. 52090020, 51722209, 51672238, 91963203, 52025026, 51525205 and U20A20238), the National Key R&D Program of China (2018YFA0703400), the NSF for Distinguished Young Scholars of Hebei Province of China (E2018203349), the Talent research project in Hebei Province (Grant No. 2020HBQZYC003) and the European Union funding via Erasmus+/DOCMASE Doctoral school (grant number "2011-0020").

**AUTHOR CONTRIBUTIONS**

Z.Z., A.V.S and Y.T. conceived the idea of this project; S.Z., B.L., Y.G. and Z.Z. prepared the samples; Z.L. and M.D. prepared the micropillars and carried out the compressive strength measurements. S.Z., Z.Z., K.L. and Y.Z. measured the XRD and Raman spectra; S.Z., K.L., Y.G., B.L., G.G. and J.H. performed hardness measurements; S.Z. and Y.G. scanned the indentations through the atomic force microscope (AFM) and scanning electron microscope (SEM); K.S, L.S. and G.Y. measured the absorption spectra; B.Z. and B.L. prepared the TEM samples using the focused ion beam (FIB) technology; W.H., A.N., Z.L. and B.L. conducted TEM and EELS characterization; Z.Z., S.Z., J.H., D.Y., B.X., Y.T., A.V.S., V.B., O.P.C., K.L., W.H., G.G., Y.L., X.Z. and B.W. analyzed the data; Z.Z., S.Z. and A.V.S. drafted the manuscript with contributions from all authors. S.Z., Z.L., K.L., J.H. and Y.G. contributed equally to this work.

**COMPETING INTERESTS**

The authors claim that the patents for current study have been granted in China (CN. 201910085279.0) and Japan (JP. 2020-009244).

**SUPPLEMENTARY DATA**

Supplementary data are available at *NSR* online.

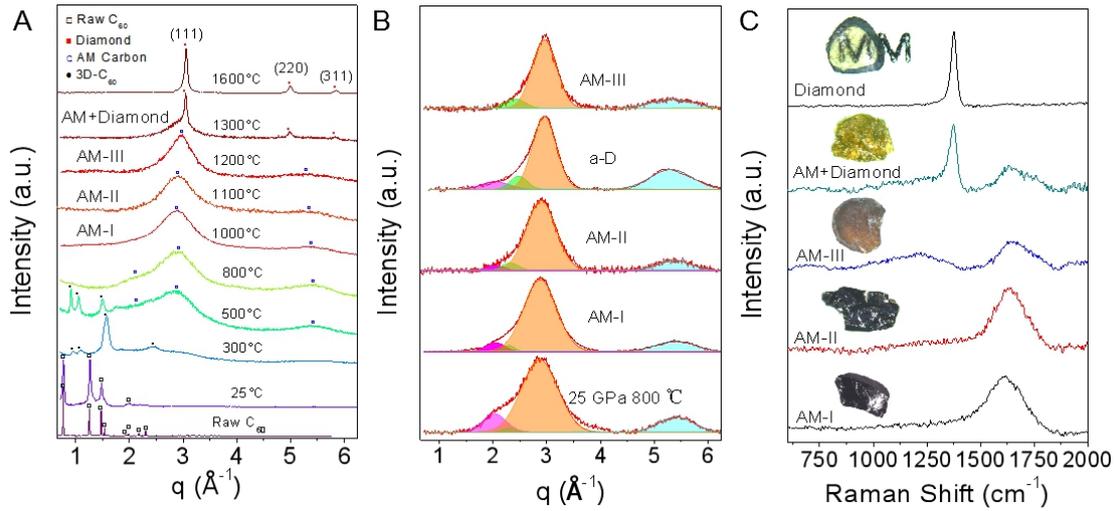

**Figure 1.** XRD patterns and Raman spectra of synthetic carbon materials collected at ambient condition. (A) XRD patterns indicating phase transition path along $C_{60}$→3D-$C_{60}$→Amorphous carbon→Diamond. AM-I, AM-II and AM-III, have one main diffraction peak at structure factor (q) of ~3.0 Å$^{-1}$ as well as another weak peak around 5.3 Å$^{-1}$, which are clearly different from previously discovered low-density amorphous carbon materials from compressing $C_{60}$ at relatively low pressures [17]. (B) Peak fitting of the XRD patterns of AM carbon materials and a-D from compressing GC [12]. The magenta, green, orange, light-blue peaks are at q=~2.0 Å$^{-1}$, ~2.4 Å$^{-1}$, ~3.0 Å$^{-1}$ and ~5.3 Å$^{-1}$, respectively. The peak at q=~2.0 Å$^{-1}$ in a-D [12], AM-I, AM-II, and amorphous carbon recovered from compressing $C_{60}$ at 25 GPa, 800 °C originates from the interlayer diffraction signal of residual graphite-like nanoclusters in the structure. This peak disappears in AM-III, demonstrating the formation of a completely different short-range ordered structure. (C) UV Raman spectra of AM-I, AM-II, AM-III, AM+Diamond composite and diamond. The insets are the optical photographs of recovered samples, displaying that AM-III is yellow-transparent and distinct from the black AM-I and M-II.

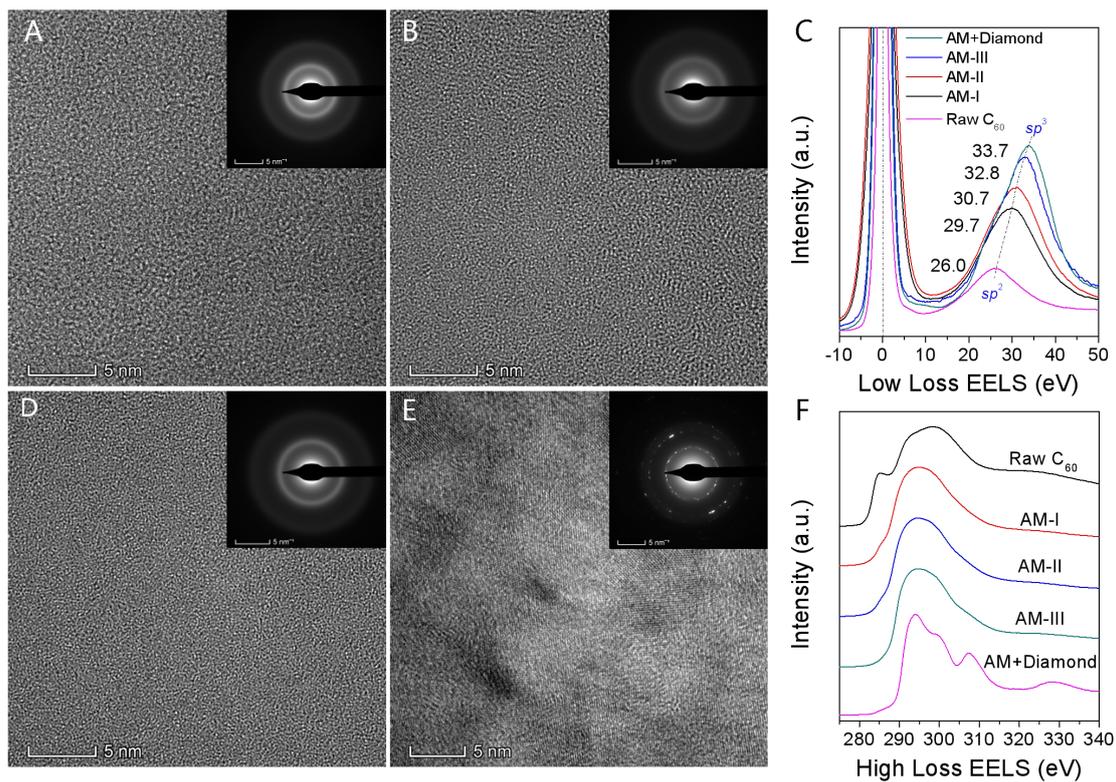

**Figure 2.** Microstructure and bonding of synthetic carbon materials. (A, B and D) HRTEM images of AM-I, AM-II and AM-III, respectively, showing their uniform disorder characteristics and gradually decreased amorphous fragment sizes. The insets: the corresponding SAED patterns, exhibit two diffuse rings near 2.1 Å and 1.2 Å. (E) HRTEM image of AM+diamond composite. The results of HRTEM and SAED pattern indicate the formation of nanocrystalline diamond. (C) Low loss EELS data show the position of plasmon peak is shifted from 26.0 eV to 33.7 eV, indicating the increase of $sp^3$ content in the samples. (F) High loss EELS show the contribution of the $sp^2$ carbon in the spectra represented by the 1s-$\pi^*$ (285 eV) transition gradually decreases with increase of the synthesis temperature (top to bottom).

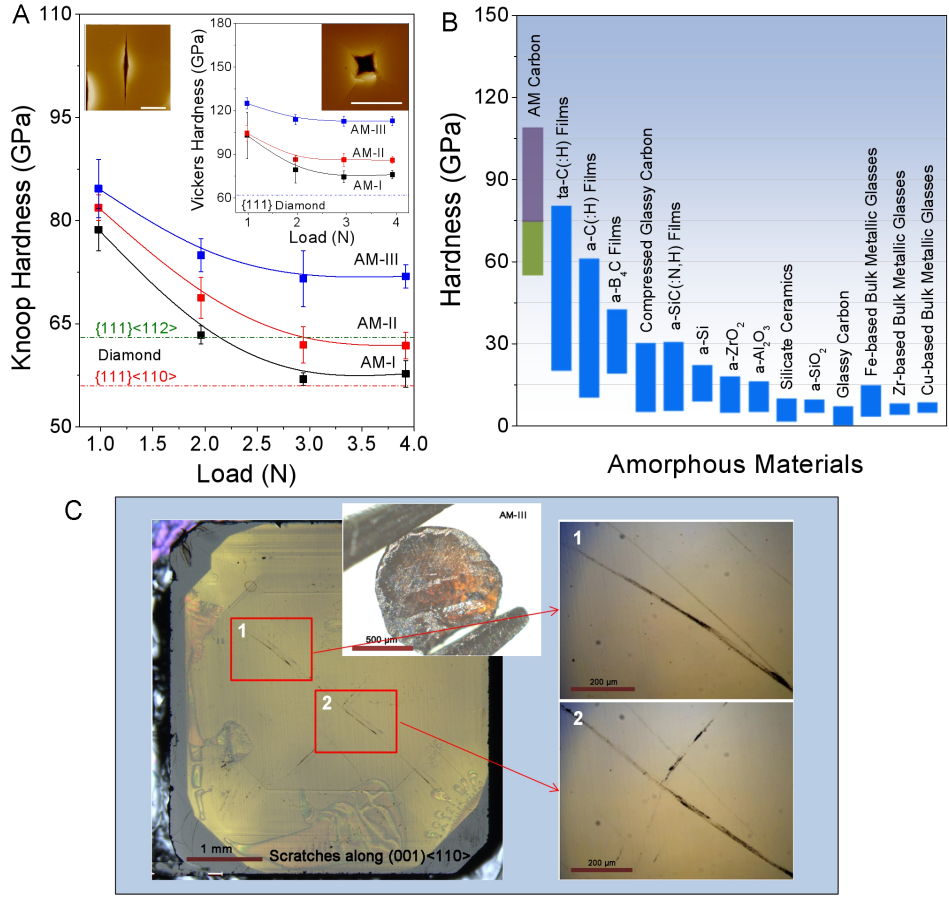

**Figure 3.** Hardness of AM carbon materials, compared with other known amorphous materials, and scratches on diamond (001) face indented by AM-III. (A) $H_K$ as a function of applied loads. Left inset: AFM image of Knoop indentation of AM-III after unloading from 3.92 N. Right inset: $H_V$ of AM carbon materials as a function of applied load and AFM image of Vickers indentation of AM-III sample after unloading from 2.94 N. The scale bars in indentation images are 10 µm. Error bars of hardness indicate s.d. (n=5). The dashed lines indicate $H_V$ and $H_K$ of (111) plane of natural diamond crystal. (B) Hardness of different amorphous materials [1, 4, 8, 11, 33]. Green and violet columns indicate $H_K$ and $H_V$ of AM carbon materials, respectively. Considering the hardness of film materials are characterized by nanoindentation hardness ($H_N$), the $H_N$ of AM carbon materials was also measured, and AM-III has high $H_N$ of 103 GPa, exceeding that (80.2 GPa) of ta-C films [8]. (C) Scratches on the (001) face of diamond by using AM-III sample displayed in the inset as an indenter (left image), indicating the ultrahard nature of AM-III. The zoomed-in right images are corresponding to the areas marked by red rectangles in the left image, displaying the scratches in more detail. (212 words)

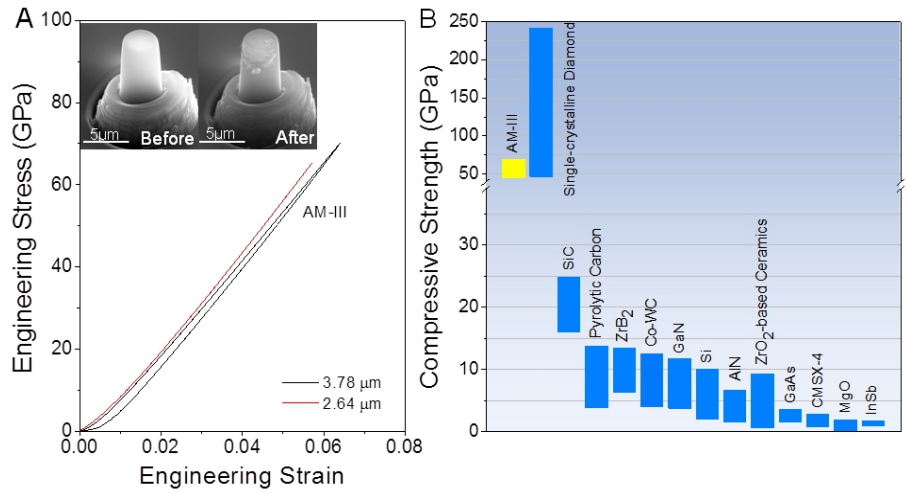

**Figure 4.** Compressive strength of AM-III compared with other known materials. (A) Engineering stress-strain curves recorded during uniaxial compressing AM-III micropillars. The insets are the SEM images of the pillar with diameter of 3.78 μm before and after compression. There was almost no size change, but some wrinkles produced on upper part of the pillar are like the shear bands formed in metallic glasses [3]. (B) Comparison of compressive strength for various materials with micron size [34–36]. The results demonstrate that AM-III is strongest amorphous materials known to date.

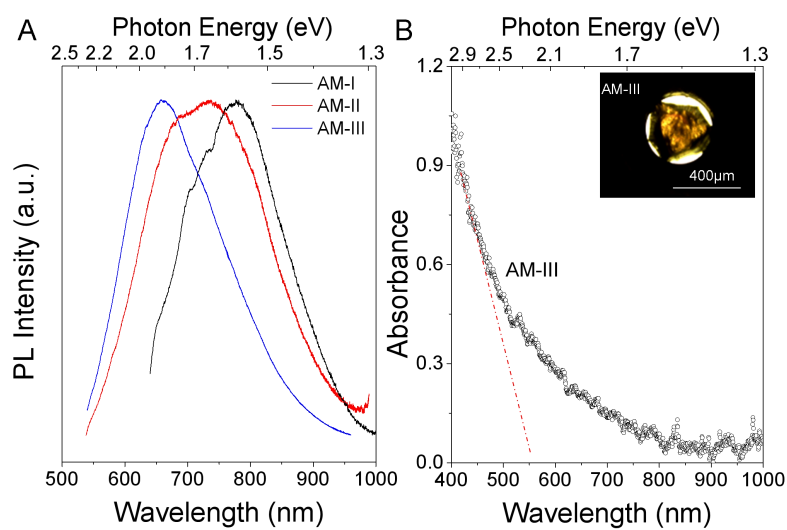

**Figure 5.** Optical properties and bandgaps of AM carbon materials. (A) PL spectra measured at ambient condition. The AM-I spectrum is excited by 633 nm laser, the AM-II and AM-III spectra are excited by 532 nm laser. The bandgaps of AM carbon materials estimated from PL spectra are between 1.5 and 2.2 eV, illustrating their semiconducting nature. (B) Absorption spectrum of AM-III. The absorption edge of AM-III is at ~570 nm, corresponding to optical bandgap value of 2.15 eV. The inset shows an optical microscope view of a piece of transparent AM-III placed inside the hole of a gasket which is mounted inside the diamond-anvil cell (DAC).

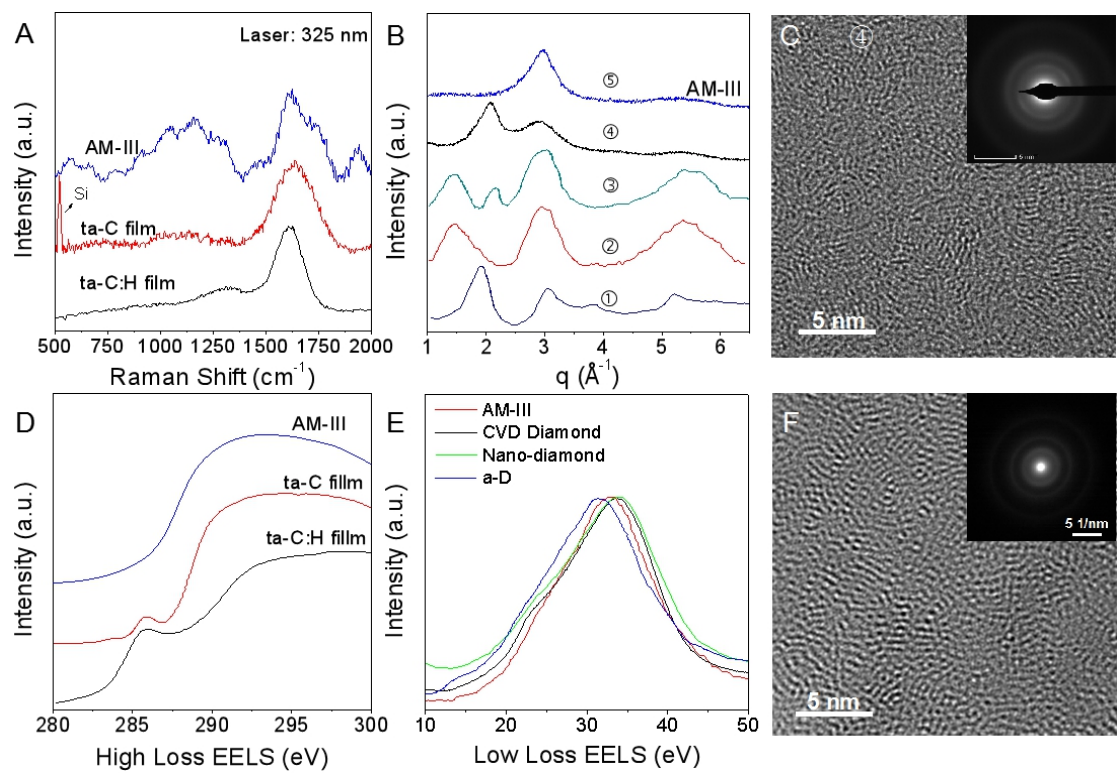

**Figure 6.** Comparison of AM-III currently discovered with other AM carbon materials. (A) Raman spectra show AM-III has an obvious T-band around 900~1300 cm$^{-1}$ compared to ta-C(:H) films [8, 27]. (B) XRD patterns of various AM carbon materials recovered from compressing $C_{60}$ at p,T conditions: ①8 GPa, 1200 °C [17]; ②12.5 GPa, 500 °C [17]; ③13.5 GPa, 1000 °C [17]; ④15 GPa, 800 °C (our data); ⑤25 GPa, 1200 °C (AM-III, our data). (C and F) HRTEM images and SAED patterns of the material synthesized at p,T condition ④ and compressed GC [11], respectively. (D) High loss EELS of AM-III and ta-C(:H) films [46, 47]. (E) Low loss EELS demonstrating the plasmon peak position in AM-III, a-D [12], CVD diamond [49], and nano-diamond [50].

# Supplementary Information

# Discovery of carbon-based strongest and hardest amorphous material

Shuangshuang Zhang[1†], Zihe Li[1†], Kun Luo[1,2†], Julong He[1†], Yufei Gao[1,2†], Alexander V. Soldatov[3,4,5], Vicente Benavides[3,6], Kaiyuan Shi[7], Anmin Nie[1], Bin Zhang[1], Wentao Hu[1], Mengdong Ma[1], Yong Liu[2], Bin Wen[1], Guoying Gao[1], Bing Liu[1], Yang Zhang[1,2], Yu Shu[1], Dongli Yu[1], Xiang-Feng Zhou[1], Zhisheng Zhao[1]*, Bo Xu[1], Lei Su[7], Guoqiang Yang[7], Olga P. Chernogorova[8], Yongjun Tian[1]*

[1]Center for High Pressure Science (CHiPS), State Key Laboratory of Metastable Materials Science and Technology, Yanshan University, Qinhuangdao, Hebei 066004, China

[2]Hebei Key Laboratory of Microstructural Material Physics, School of Science, Yanshan University, Qinhuangdao 066004, China

[3]Department of Engineering Sciences and Mathematics, Luleå University of Technology, SE-97187 Luleå, Sweden

[4]Department of Physics, Harvard University, Cambridge, MA 02138, USA

[5]Center for High Pressure Science and Technology Advanced Research, Shanghai 201203, China

[6]Department of Materials Science, Saarland University, Campus D3.3, D-66123, Saarbrücken, Germany

[7]Key Laboratory of Photochemistry, Institute of Chemistry, University of Chinese Academy of Sciences, Chinese Academy of Sciences, Beijing, 100190, China

[8]Baikov Institute of Metallurgy and Materials Science, Moscow 119334, Russia

* Corresponding authors: zzhao@ysu.edu.cn (Z.Z.) or fhcl@ysu.edu.cn (Y.T.). †These authors contributed equally to this work.

## Table of Contents

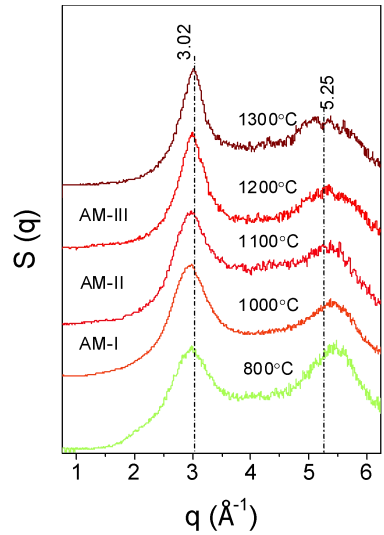

**Figure 1.** Structure factors S(q) of the carbon materials quenched from synthesis pressure of 25 GPa and various temperatures. Two broad diffraction peaks are visible at positions of ~3.0 and 5.3 Å$^{-1}$, respectively. With the synthesis temperature increase, the first and second peak shift to higher- and lower q, gradually approaching the (111) and (220) reflections position of diamond at q=3.05 and 4.98 Å$^{-1}$, respectively. The dashed lines give an indication of the peak shifts.

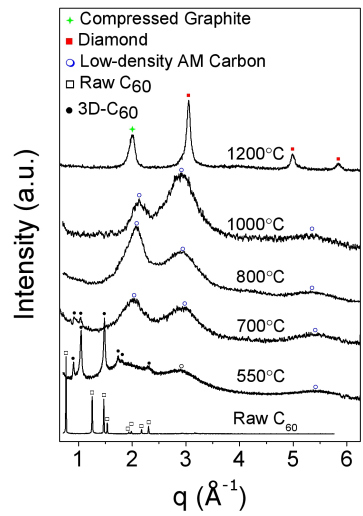

**Figure 2.** XRD patterns of carbon materials synthesized at 15 GPa and different temperatures from collapsed $C_{60}$ fullerenes collected after subsequent samples quenching to ambient conditions. With the increased synthesis temperatures, the resulting phase transition path is $C_{60}$→3D-$C_{60}$→ Amorphous carbon→Diamond/compressed graphite composite. At this synthesis condition, the AM carbon materials recovered from 700-1000 °C have two main diffraction peaks at around 2.0 Å$^{-1}$ and 2.9 Å$^{-1}$, as well as one minor peak at about 5.3 Å$^{-1}$. In contrast, the AM carbon materials recovered from 25 GPa and 1000-1200 °C have only two diffraction peaks, at around 3.0 Å$^{-1}$ and 5.3 Å$^{-1}$. The diffraction peak centered around 2.0 Å$^{-1}$ corresponds to the graphite interlayer-like distance of ~3.1 Å.

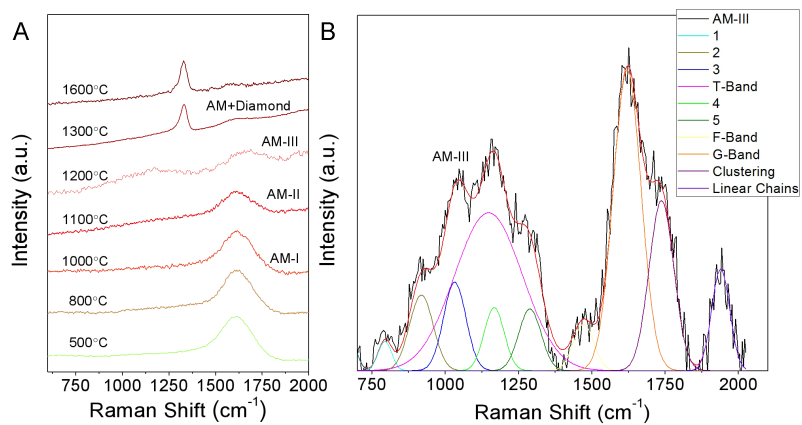

**Figure 3.** UV Raman spectra of the carbon materials quenched from synthesis pressure of 25 GPa and various temperatures. (A) The fluorescence background was not removed. Above 1300 °C, the Raman peak of diamond appeared at 1330 cm$^{-1}$. Below that synthesis temperature, all the spectra exhibit characteristic of *sp*$^2$ carbon broad G-band located at about 1600 cm$^{-1}$. An additional broad peak appears at around 900~1300 cm$^{-1}$ in AM-III, which is commonly known as T-band indicating the high *sp*$^3$ carbons fraction in this material. (B) Decomposition of the UV Raman spectrum of the AM-III and peaks assignment. Peaks 1 to 5: UV Raman spectroscopy of hydrocarbons [1] reveals similar vibrations that can be related to different configuration of fused aromatic rings. In case of AM-III it is likely that the aromatic rings are linked to each other (fused) via *sp*$^3$ carbons and randomly oriented in the material; F-Band: this vibration is related to pentagonal rings (analogous to Ag(2) mode in C$_{60}$). F-band was observed in fullerene-like disordered carbon systems [2], like glassy carbon [3], fullerene-like amorphous carbon thin films [4, 5] and nano-clustered graphene [6]; G-Band peak position reflects a mixture of linear chains and fused rings fragments [2]; "C-clustering" peak at 1740 cm$^{-1}$ corresponds to tiny graphene clusters residues in highly disordered *sp*$^3$/*sp*$^2$ carbon systems, for example, carbon nanodots [7]; peak at 1940 cm$^{-1}$ is tentatively assigned to short linear carbon chains [2, 8].

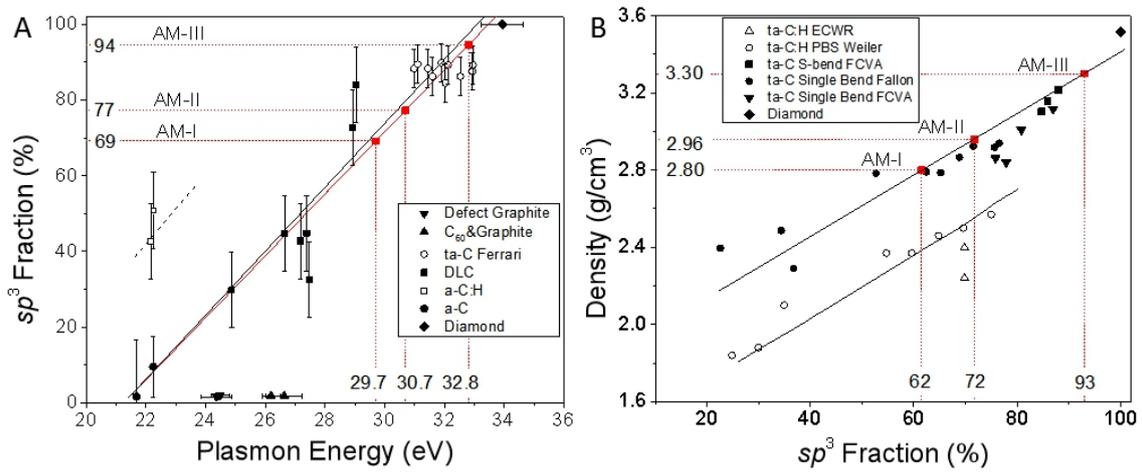

**Figure 4.** (A) Variation of $sp^3$ fraction vs. plasmon energy for various carbon materials [11]; the red line is the result of linear re-fitting, based on the combined data after adding the Ferrari' data [12]. (B) Variation of density vs. $sp^3$ fraction for various carbon materials [13]. The $sp^3$ fraction in AM-I, AM-II, and AM-III was estimated according to above linear relationship (see red dots).

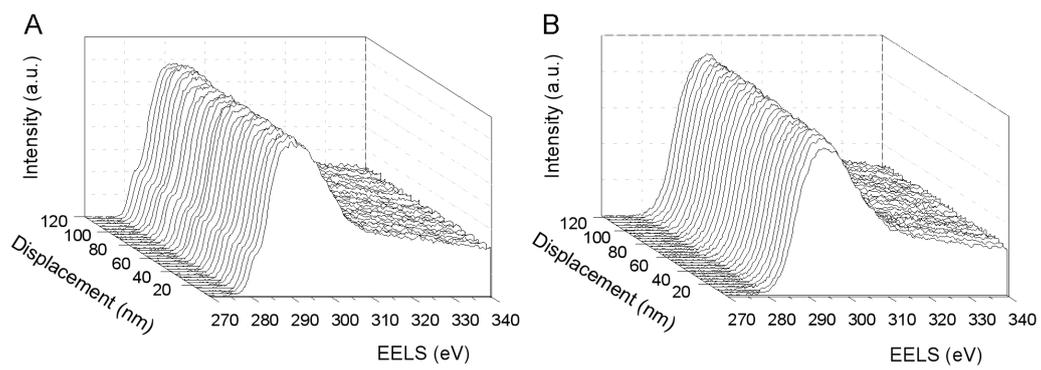

**Figure 5.** EELS line scans conducted in STEM mode along ~120 nm long lines with 1 nm step and energy resolution of 0.6 eV in a randomly selected regions of the AM-I (A) and AM-III (B) samples, respectively, showing the bonding homogeneity in their microstructures.

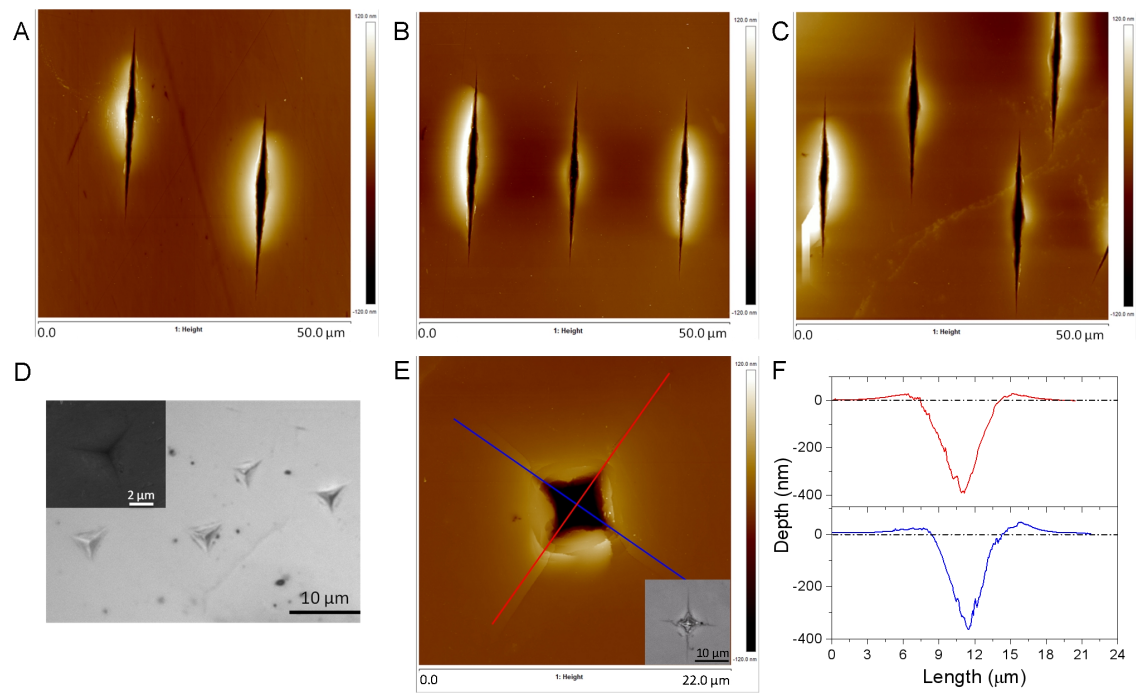

**Figure 6.** Images of Knoop, Vickers and Berkovich indentations. (A, B and C) AFM images of Knoop indentations on surfaces of AM-I, AM-II and AM-III phases, respectively. In all cases, the applied load was 3.92 N. (D) Optical and SEM images of AM-III phase surface after indentation at a load of 0.98 N with Berkovich-type pyramid probe. (E) AFM image of a residual indentation by the Vickers probe on AM-III surface at a load of 2.94 N. The inset shows corresponding optical photograph of the indentation. (F) AFM scan of the indentation profile in e along the diagonals.

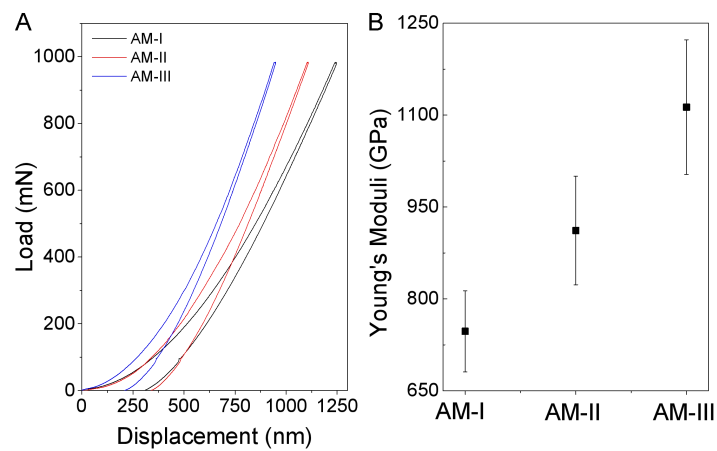

**Figure 7.** Nanoindentation hardness and Young's moduli of AM-I, AM-II and AM-III. (A) Loading/unloading displacement curves during indentation measurement. The derived hardness ($H_N$) at a peak load of 0.98 N are 76±3.4, 90±7.9, and 103±2.3 GPa, respectively, which are comparable to the hardness values determined by Vickers method. (B) Young's moduli ($E$) of the AM carbon materials. By assuming Poisson's ratio of 0.2, the estimated $E$ of AM-I, AM-II and AM-III are 747±66, 912±89, 1113±110 GPa, respectively.

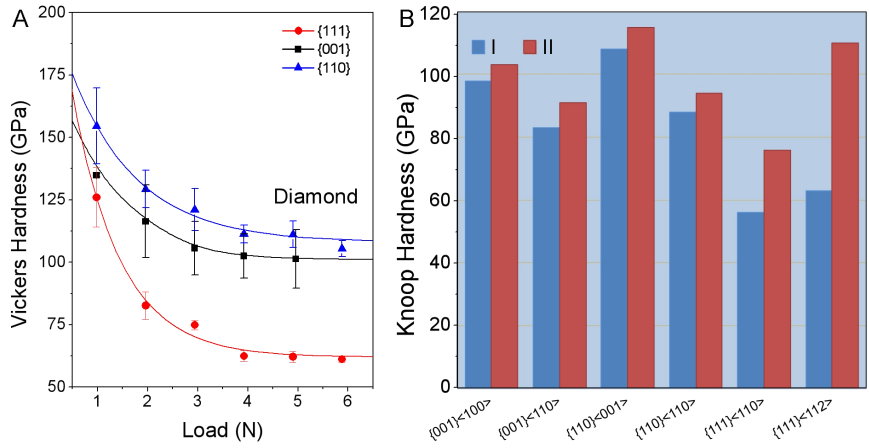

**Figure 8.** Vickers ($H_V$) and Knoop ($H_K$) hardness of different crystal faces of single crystalline diamond. (A) $H_V$ as a function of applied load. The asymptotic $H_V$ values of {111} and {110} faces of natural diamond are 62 and 111 GPa, respectively [14]. In this work we determined the asymptotic $H_V$ of {001} face of synthetic diamond at 103 GPa. (B) $H_K$ of natural diamond along different crystallographic directions. The data in the figure are from the literature [15] (I: Type Ia natural diamond, and II: Type IIa natural diamond).

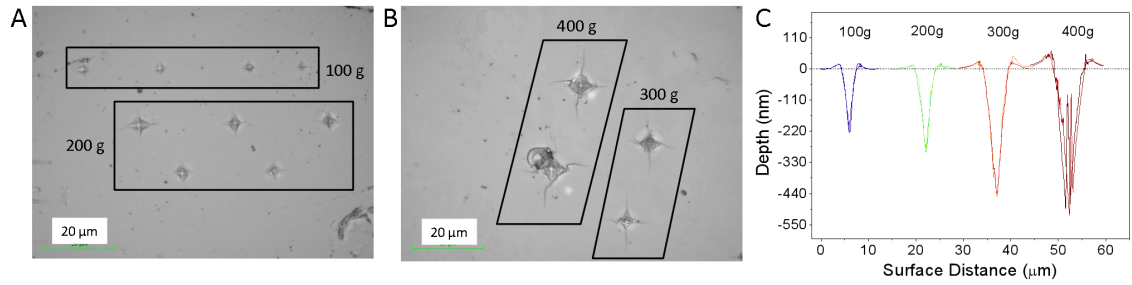

**Figure 9.** Vickers indentation morphologies of AM-III after unloading from different loads (A and B) and the corresponding indentation profiles along the diagonals scanned by AFM (C). At small loads, there is no obvious indentation cracks, indicating the dominant plastic deformation. At the large loads, the radial and lateral cracks as well as peeling zones can be found around the indentations, demonstrating the plastic-to-brittle transition [16]. For all the loads, the displaced material flows up around the indenter to form a raised pile-up, indicating the occurrence of plastic flow in these cases.

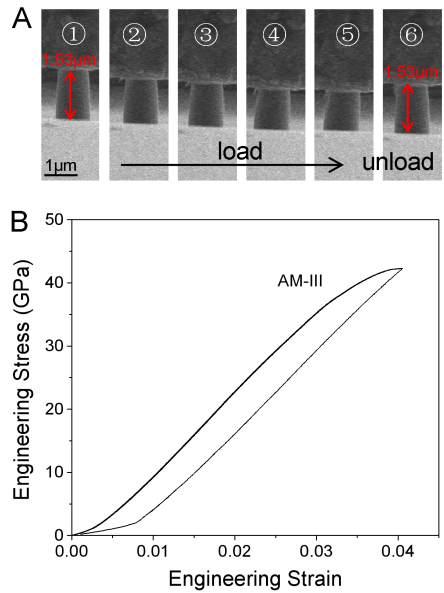

**Figure 10.** *In-situ* compression/decompression testing of an AM-III micron-sized pillar. (A) *In-situ* SEM images exhibit the pillar height change during compression (①-⑤) and after decompression ⑥. The micron-sized pillar with a top diameter of 0.88 μm was shortened during compression, and the deformation was completely recovered after unloading. (B) Engineering stress-strain curve. Notably, the compression curve shows nonlinearity at the maximum load, likely due to tilting and bending of the pillar.

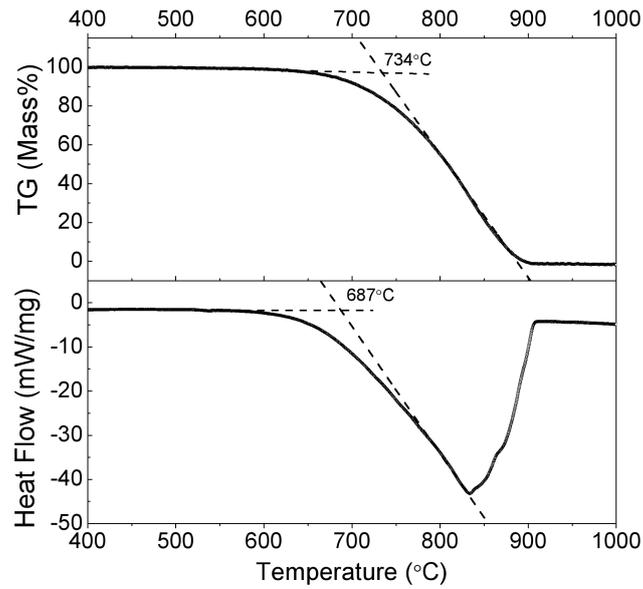

**Figure 11.** Thermogravimetric analysis (TGA) (top panel) and differential scanning calorimetry (DSC) heat flow data (bottom panel) collected from AM-III in air. The oxidation onset temperatures were determined at 734 °C and 687 °C, from TGA and DSC data, respectively. The thermal stability of AM-III is comparable to that of single crystalline diamond [14].